\documentclass[a4paper]{jpconf}
\usepackage{graphicx}
\usepackage{amsmath,amssymb}
\usepackage{epsfig}
\usepackage{dcolumn}
\usepackage{bm}
\usepackage{theorem}
\usepackage{amsmath,amssymb}

\newcommand{\qed}{\nobreak \ifvmode \relax \else
\ifdim\lastskip<1.5em \hskip-\lastskip \hskip1.5em plus0em
minus0.5em \fi \nobreak \vrule height0.75em width0.5em
depth0.25em\fi}

\input{epsf}

\newcommand{\bmp}{\begin{minipage}}
\newcommand{\emp}{\end{minipage}}
\newcommand{\bc}{\begin{center}}
\newcommand{\ec}{\end{center}}
\newcommand{\beq}{\begin{equation}}
\newcommand{\eneq}{\end{equation}}

\begin{document}
\title{Molecular architectures based on $\pi$-conjugated block copolymers for global quantum computation }
\author{C A Mujica Martinez$^{1,3}$, J C Arce$^1$, J H Reina$^2$ and M Thorwart$^{3,4}$}
\address{$^1$Universidad del Valle, Departamento de Qu\'imica, A. A. 25360, Cali, Colombia}
\address{$^2$Universidad del Valle, Departamento de F\'isica, A. A. 25360, Cali, Colombia}
\address{$^3$Institut f\"ur 
Theoretische Physik IV, Heinrich-Heine-Universit\"at D\"usseldorf,
40225  D\"usseldorf, Germany}
\address{$^4$Freiburg Institute for Advanced Studies (FRIAS), Universit\"at Freiburg,
79104 Freiburg, Germany}
\ead{camujica, jularce, and  jhreina@univalle.edu.co
}
%
\begin{abstract}
We propose a molecular setup for the physical implementation of a  barrier {\it global} quantum computation scheme based on the electron-doped $\pi$-conjugated copolymer architecture of nine blocks PPP-PDA-PPP-PA-(CCH-acene)-PA-PPP-PDA-PPP (where each block is an oligomer). 
The physical carriers of information are electrons coupled through the Coulomb interaction, and the building block of the computing architecture is composed by three adjacent  qubit systems in a quasi-linear arrangement,  each of them allowing qubit storage, but with the central qubit exhibiting a third accessible state of electronic energy far away from that of the qubits' transition energy. The third state is reached from one of the computational states by means of an on-resonance coherent laser field, and acts as a barrier mechanism for the direct control of qubit entanglement. 
Initial estimations of the spontaneous emission decay rates associated to the energy level structure allow us to compute a damping rate of order $10^{-7}$ s, which suggest a not so strong coupling to the environment.  Our results offer an all-optical,  scalable, proposal for global quantum computing based on semiconducting $\pi$-conjugated polymers.
\end{abstract}
\noindent
To appear  in {\jpcs}

\section{Introduction}\label{sec:01}
The quantum processing of information has caused great impact from the standpoint of its potential uses, to the point of being considered a `second quantum revolution' \cite{1}. One of the advantages of these technologies is their capability of encrypting information with extreme security, avoiding the interception of communications between two remote points \cite{2}. Another advantage is the possibility of achieving quantum computation, which exploits the interference principle and the existence of non-local correlations between interacting quantum objects, which have become fundamental physical resources in the development of quantum information protocols. This would allow a quantum computer to carry out tasks that are intractable for a classical computer (e.g. the factorization of a large prime number \cite{3} or the evaluation of a whole domain of a function in a single computational step \cite{2,4}); nevertheless, the experimental challenge implied by the construction of this kind of devices stems from the difficulty of {\it locally} controlling the quantum dynamics of the quantum bits (qubits) and the decoherence caused by the interactions with the environment \cite{5,5a,6}.

A qubit is a quantum system in which the Boolean states (classical bits) $ 0 $ and $ 1 $ are represented by a pair of orthonormal quantum states identified as $|0\rangle $  and $ |1\rangle$; the two states form a computational base, and any other state of the qubit can be written as a superposition $\alpha|0\rangle + \beta|1\rangle$, with $\alpha $ and $\beta$ complex numbers in general, such that $ |\alpha|^2 + |\beta|^2 = 1 $. Physically, a qubit can be a microscopic system like an atom \cite{7,8}, a nuclear spin \cite{9}, a polarized photon \cite{10}, or more complex structures like quantum dots \cite{11,16a,16b,16c}, molecular arrays \cite{12,13}, and photosynthetic biomolecular systems \cite{14}, among many others. 

Quantum dots are advantageous due to the existing industry for their nanofabrication and the ease of incorporating them into current opto-electronic devices \cite{15}; besides, quantum computation schemes using quantum dots have been reported \cite{11,16a,16b,16c,16,17,18}. These quantum dots can be constructed from inorganic or organic semiconductors, the latter being of special interest due to previous studies of such \cite{19}, where the quantum dots are really organic heterostructures (block copolymers). They are easier to construct than the inorganic systems, since they do not require expensive pieces of equipment as required for molecular beam epitaxy or metal-organic chemical vapour deposition. In principle, it is possible to construct an unlimited variety of organic heterostructures, since the interface between the materials is a chemical carbon-carbon bond, in contrast to inorganic ones, where it is required that the materials exhibit similar lattice constants to avoid interfacial stress \cite{20}, which  notably limits the variety of heterostructures that can be synthesised.

Nowadays, many theoretical schemes for quantum computation in the solid state can be found. Universal quantum computation requires controlling the dynamics of individual qubits and the interactions between them, the latter being the main cause of decoherence (loss of coherence between quantum superpositions) due to the introduction of control mechanisms external to the system and the ubiquituous coupling to the environment. The use of electromagnetic pulses as a control mechanism does not cause severe decoherence, which has motivated the proposal of all-optical quantum computation schemes \cite{21,22,23}.
\begin{figure}[h]
\begin{center}
\includegraphics[width=0.465\textwidth]{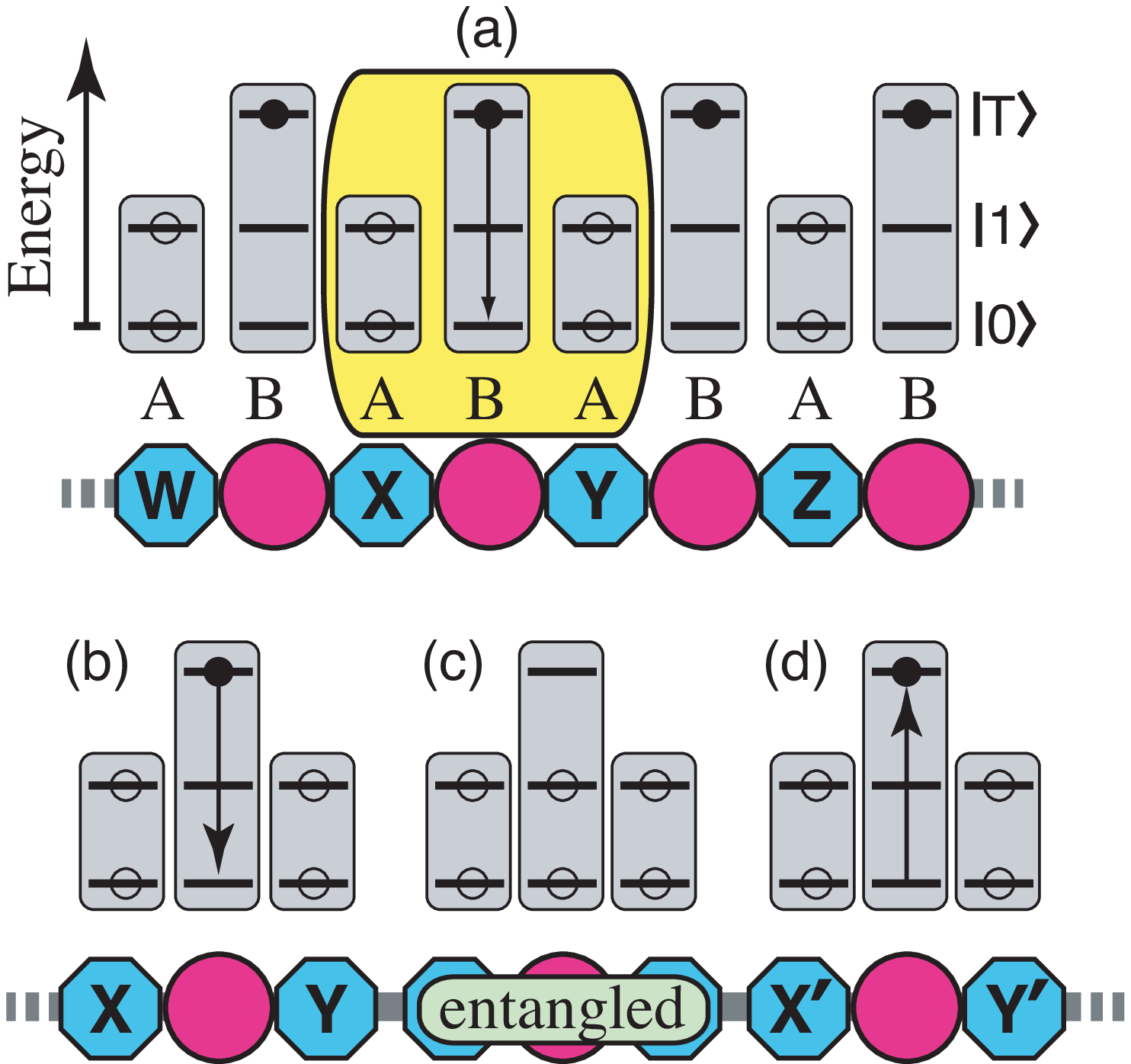} 
\includegraphics[width=0.52\textwidth]{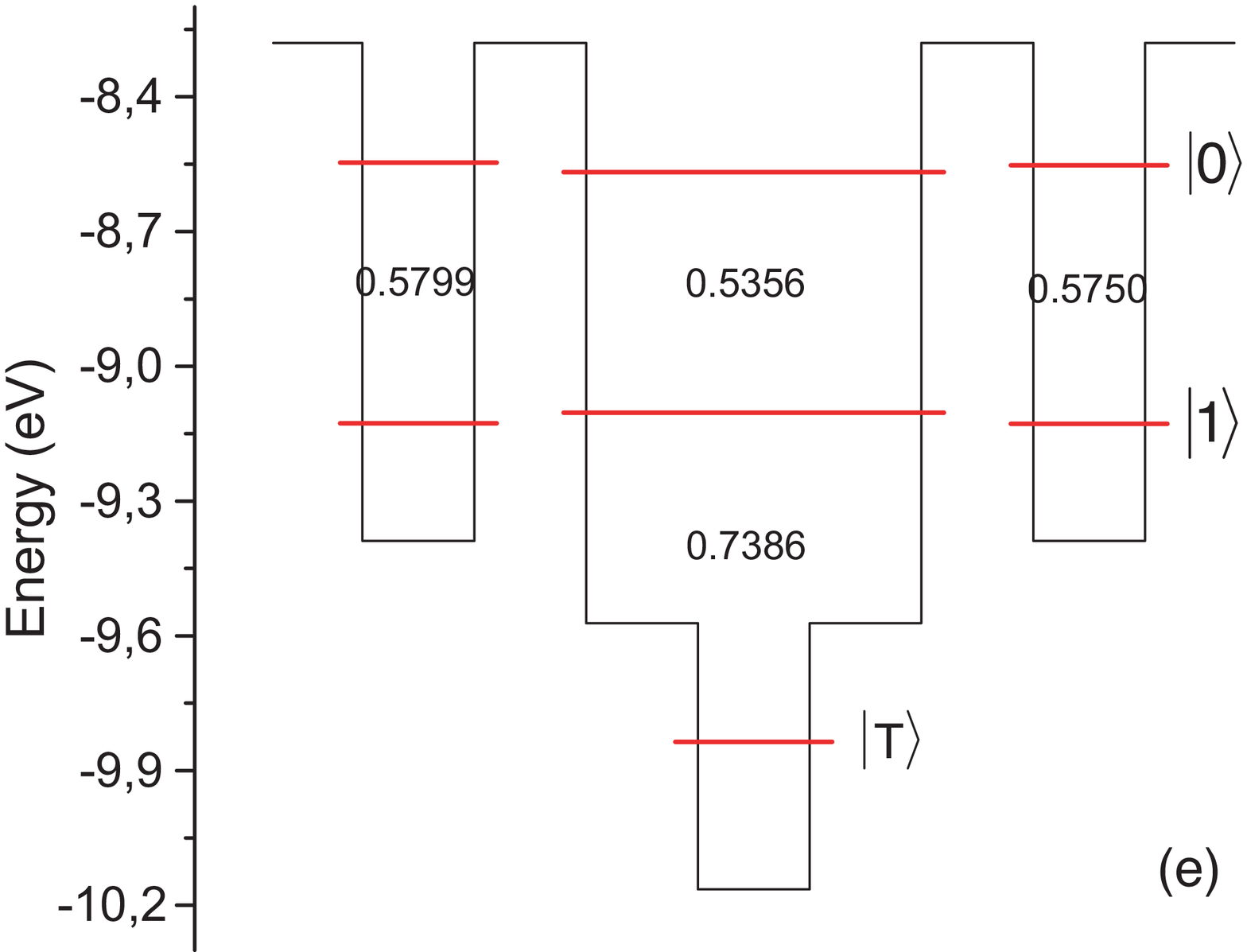}
\end{center}
\caption{ \label{fig1}  (a) Qubits ($A$ species) coupled through a barrier ($B$ species). (b) The qubits are denoted in the computational basis $\{|0\rangle ,|1\rangle\}$, and the central system possesses a third accessible state $ |T\rangle $. The barrier is ``removed" and all the qubits interact so that they (c) get entangled. (d) The barrier is raised again, once the lateral qubits $X'$ and $Y'$ are decoupled from the central qubit.  (e) Conduction-band profile and monoelectronic levels of the proposed molecular architecture, calculated by means of the extended H\"uckel method. Energies are given in eV.
} 
\end{figure}

\section{Model}

Various control schemes have been proposed to control decoherence, among which the spin-barrier one stands out \cite{6,22,23}, which employs two types of two-level systems in a linear arrangement: one type carries the information (qubits) and the other type acts as passive barriers of global control between qubits. In this work we propose the implementation of a model based on the following architecture \cite{23}: Fig. 1(a) shows a linear arrangement of qubits (�A� species) and control units (�B� species) which possess a third additional level $|T\rangle$ of electronic energy well off resonance with the levels of the computational qubits ($\{|0\rangle ,|1\rangle\}$). The linear arrangement is illustrated schematically in the lower part of Fig. 1(a), where the circles represent the �barriers� and the octagons the computational qubits ($W,X,Y,Z$).

Fig. 1(b) shows the basic block $ABA$ of Fig. 1(a) at work. In it, the barrier is prepared in the state $|T\rangle$ by the excitation with an external laser, which precludes the interaction between the lateral qubits $X$ and $Y$ (Fig. 1(b)), since $|T\rangle$ is out of resonance with $X$ and $Y$. To carry out a logical gate, the central system decays to the subspace $\{|0\rangle ,|1\rangle\}$ (Fig. 1(c)), and the three systems become entangled under free evolution, which allows the quantum gates to be carried out (Fig. 1(c)). To �turn off� the interaction, the laser is applied again in such a way that the state $|T\rangle$ is populated again, restoring the barrier, as shown in Fig. 1(d), where a product state of the qubits has been recovered, now represented as ${|X'\rangle \otimes |Y'\rangle}$.

\section{Molecular architecture based on $\pi$-conjugated multi-block copolymers }

In this work, we implement the model illustrated above employing negatively doped $\pi$-conjugated multi-block copolymers \cite{19}, where the carriers of the physical information are electrons interacting through the Coulomb force. This architecture has the advantage of permitting a strong coupling between qubits, since it is molecular \cite{23}, interacting weakly with phonons, since it is quasi-linear \cite{19}, and being scalable. By adjusting the length of the wells and barriers, the nine-block architecture 8 PPP-7 PDA-8 PPP-3 PA-7 (CCH acene)-3 PA-8 PPP-7 PDA-8 PPP was devised, where PPP, PDA, PA and CCH acene refer to poly-(\emph{p}-phenylene), poly-diacetylene, \emph{trans}-polyacetylene and poly-acene substituted with ethynyl (CCH) groups.

According to the envelope molecular orbital theory \cite{19}, the conduction band of the nine-blocks can be visualised as the multiple well illustrated in Fig. 1(e), where the mono-electronic levels localised in each well and their energy differences are indicated. Such levels were determined employing the semiempirical extended H\"uckel method \cite{24}, which includes all the valence electrons and takes into account the overlap between all atomic basis functions.

\section{Results}

To visualise compactly the shape of the molecular orbitals in one dimension, a localisation parameter on the $j$-th atom is defined as \cite{19}
\begin{equation}
 L_j^{(i)} = \sum_n |c_{jn}^{(i)}|^2 ,
\end{equation}
where $n$ labels the atomic orbitals that enter the expression of the $i$-th H\"uckel molecular orbital in terms of a linear combination of atomic orbitals
\begin{equation}
 |\psi_i\rangle = \sum_r c_{r}^i|\phi_r\rangle .
\end{equation}
Since $c_{jn}^{(i)}$ is the contribution of the $n$-th basis AO centered on the $j$-th atom to the $i$-th MO, $L_j^{(i)}$ provides a measure of the electron density at that site of the molecule. Since the value of this parameter is a function of the position along the chain, specified by $j$, the set of $L_j^{(i)}$ reveals the spatial shape of the MO. In Figs. 2(a)-(h) the localisation parameters are shown for the MOs of the conduction band of the heterostructure.

As Fig. 2 shows, the MOs are localised within one of the wells, which justifies the assignment of each eigenstate to a certain spatial region of the molecule, as indicated in Fig. 1(e).
When each of the three wells is doped with an electron, the states $ |0\rangle $ and $ |1\rangle $ of the lateral wells act as qubits, whereas the state of the central well permits activation and deactivation of the resonant coupling, and therefore the entanglement between them through stimulated optical transitions $ |0\rangle \longleftrightarrow |T\rangle $. It must be noted that the ordering of the energy levels is the inverse of the original scheme of Fig. 1(a).
\begin{figure}[h]
\begin{center}
\includegraphics[width=0.49\textwidth]{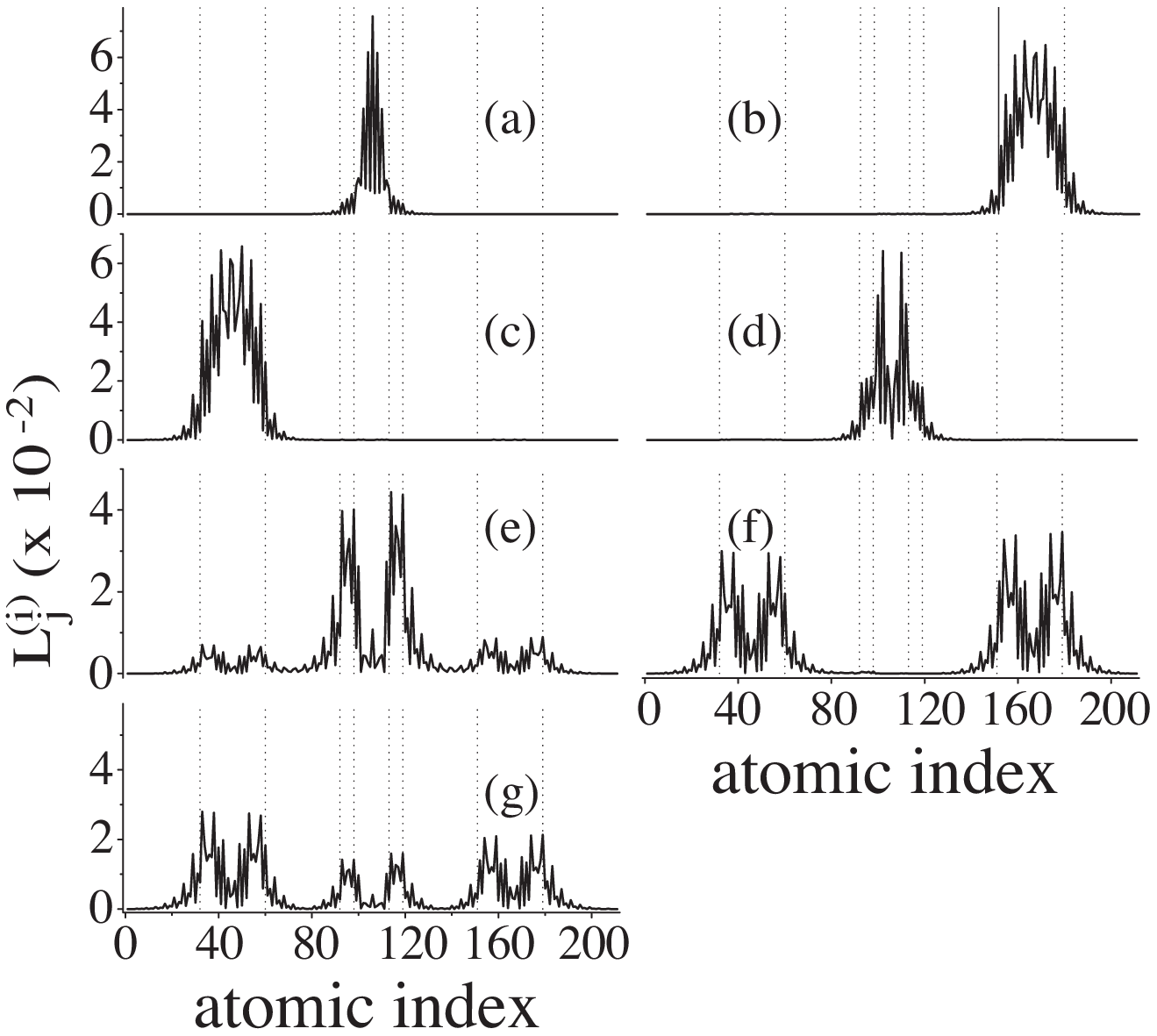} \,
\includegraphics[width=0.49\textwidth]{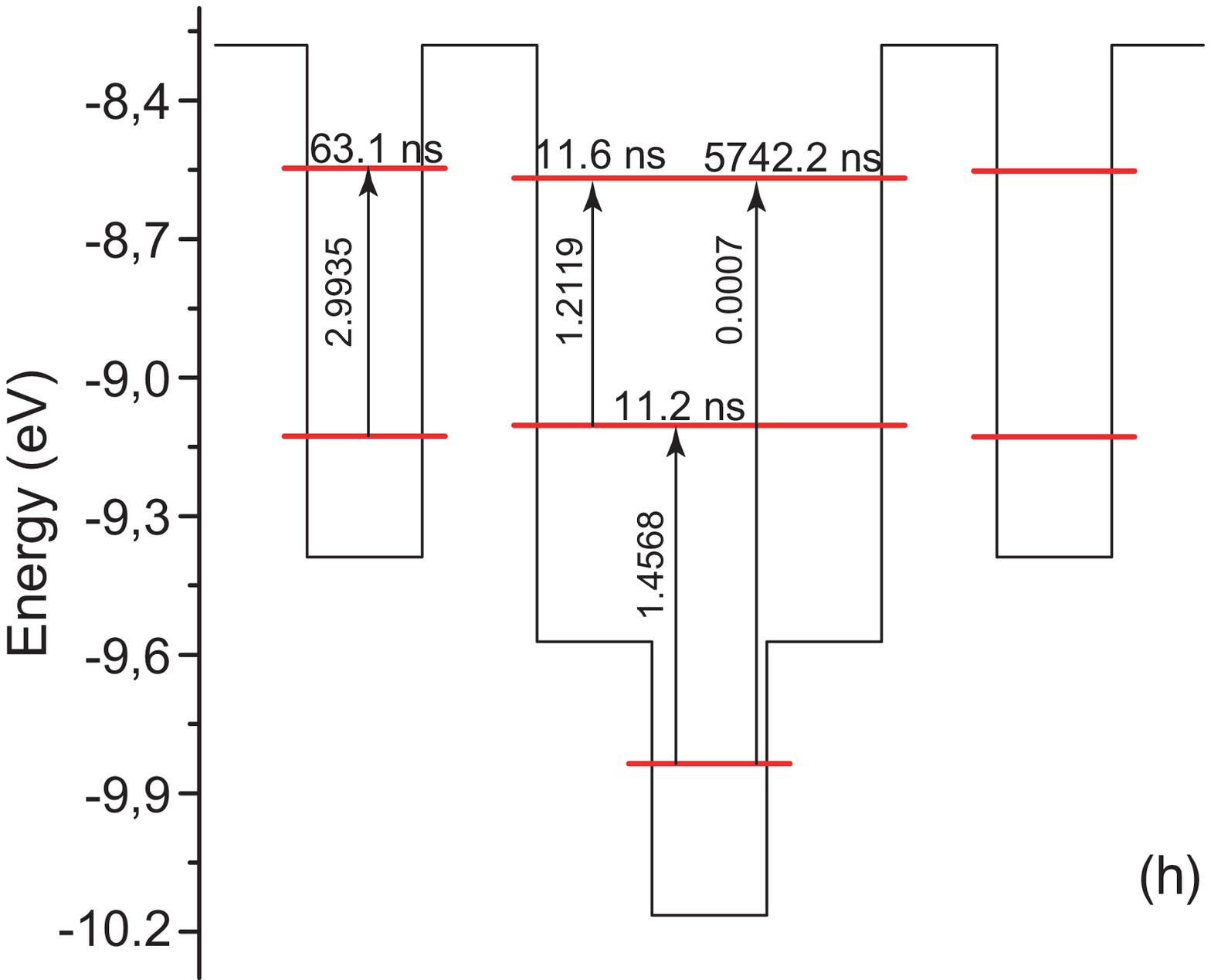}
\end{center}
\caption{ \label{fig3} Behaviour of the localisation parameter for the MOs of the conduction band of the heterostructure. (a) LUMO, (b) LUMO+1, (c) LUMO+2, (d) LUMO+3, (e) LUMO+4, (f) LUMO+5, (g) LUMO+6. The dashed lines indicate the position of the barriers of each well. (h) Mean lifetimes (in ns) and oscillator strengths for the molecular architecture, calculated by means of the semiempirical method ZINDO/S \cite{26}.
} 
\end{figure}

From a calculation of the oscillator strengths for the various possible transitions, we obtain  the Einstein coefficients for spontaneous emission, which, in turn, are related to the mean lifetimes of the excited states by $ \gamma = \tau^{-1}$ \cite{25}. These lifetimes provide information about the relaxation times of the system and give an indication about the plausibility of the architecture proposed. The calculated values of $\tau$ are reported in Fig. 2(h).
Assuming a simple exponential decay of the populations of the form $ e^{-\gamma t}$, we relate $\gamma $  with the damping parameter $K$: $\gamma = K\Delta\varepsilon $, where  $\Delta\varepsilon $ is the energy difference between the computational levels considered. For the designed heterostructure proposed here, $ K$ is of the order of $ 10^{-7} $. Although there is some small degree of energy mismatch in the designed heterostructure, this can be minimized by refining the quantum-chemistry calculations, and the main source of decoherence would come from electron-phonon coupling effects.

\section{Conclusions}
The results obtained indicate that the proposed architecture is a plausible system for the physical implementation of global quantum computation schemes. It fulfills the requirement of scalability, and, in addition, the control of the dynamics can be carried out using pulses of the order of picoseconds, which would allow the efficient realisation of one and two qubits unitary transformations, the elementary building blocks for performing a conditional quantum dynamics. 
 The initial estimations made for the heterostructure relaxation rates indicate that the system is not very susceptible to prompt environment-driven decoherence, which is a good indicator that the quantum coherent control proposed in this work is feasible. 

\ack 
We acknowledge financial support from Colciencias under contract 1106-45-221296, the scientific exchange program PROCOL (DAAD-Colciencias), and by the Excellence Initiative of the German Federal and State Governments.


\section*{References}
\begin{thebibliography}{10}

\bibitem{1} Dowling J P and Milburn G J 2003 \emph{Phil. Trans. R. Soc. Lond.} A \textbf{361} 1655

\bibitem{2} Nielsen M A and Chuang I L  2000 \textit{Quantum Computation and Quantum Information} (Cambridge: Cambridge University Press)

\bibitem{3} Shor P W 1997 \emph{SIAM J. Comput.} \textbf{26} 1484

\bibitem{4} Deutsch D and Jozsa R 1992 \emph{Proc. R. Soc. London} A \textbf{439} 553

\bibitem{5} Palma M, Suominen K-A and Ekert A K 1996 \emph{Proc.: Math., Phys. Eng. Sci.} \textbf{452} 567

\bibitem{5a} Reina J H, Quiroga L and Johnson N F 2002 \textit{Phys. Rev.} A {\textbf 65} 032326

\bibitem{6} Benjamin S C 2004 \emph{New J. Phys.} \textbf{6} 61

\bibitem{7} Cirac J I and Zoller P 1995 \emph{Phys. Rev. Lett.} \textbf{74} 4091

\bibitem{8} Briegel  H-J \textit{et al.} 2000 \emph{J. Mod. Opt.} \textbf{47} 415

\bibitem{9} Jones J A and Mosca M 1998 \emph{J. Chem. Phys.} \textbf{109} 1648

\bibitem{10} O'Brien J L 2007 \emph{Science} \textbf{318} 1567

\bibitem{11} Loss D and DiVincenzo D P 1998 \emph{Phys. Rev.} A \textbf{57} 120

\bibitem{16a} Reina J H, Quiroga L and Johnson N F 2000 \textit{Phys. Rev.} A {\textbf 62} 012305

\bibitem{16b} Lovett B, Reina J H, Nazir A and Briggs A 2003 \textit{Phys. Rev.} B {\textbf 68} 205319

\bibitem{16c}  Thorwart M and H\"anggi P 2002 {\it Phys. Rev.} A {\bf 65} 012309

\bibitem{12} Gershenfeld N and Chuang I L 1998 \emph{Sci. Am.} \textbf{278} 66

\bibitem{13} Tesch C M and Vivie-Riedle R 2002 \emph{Phys. Rev. Lett.} \textbf{89} 157901

\bibitem{14} Thorwart M, Eckel J, Reina J H and Weiss S (\textit{Preprint} cond-mat.mes-hall/0808.2906)

\bibitem{15} Reed M A 1993 \emph{Sci. Am.} \textbf{268} 118

\bibitem{16} Chen G \textit{et al.} 2000 \emph{Science} \textbf{289} 1906

\bibitem{17} DiVincenzo D P 2005 \emph{Science} \textbf{309} 2173

\bibitem{18} Berezovsky J \textit{et al.} 2008 \emph{Science} \textbf{320} 349

\bibitem{19}  Mujica Martinez C A 2006 \emph{Effective mass molecular orbital theory applied to quasi-one-dimensional polyacenes and their heterostructures}, undergraduate thesis, Universidad del Valle 

\bibitem{20} Yu P T and Cardona M 1999 \emph{Fundamentals of Semiconductors: Physics and Material Properties} 2nd ed (Berlin: Springer-Verlag)

\bibitem{21} Calarco T \textit{et al.} 2003 \emph{Phys. Rev.} A \textbf{68} 012310

\bibitem{22} Benjamin S C and Bose S 2003 \emph{Phys. Rev. Lett.} \textbf{90} 247901

\bibitem{23} Benjamin S C , Lovett B W and Reina J H 2004 \emph{Phys. Rev.} A \textbf{70} 060305R

\bibitem{24} Hoffmann R 1963 \emph{J. Chem. Phys.} \textbf{39} 1397

\bibitem{25} McHale J L 1999 \emph{Molecular Spectroscopy} (New York: Prentice Hall)

\bibitem{26} Ridley J E and Zerner M C 1973 \emph{Theo. Chem. Acta} \textbf{32} 111


\end{thebibliography}
\end{document}